# Astrobiology

## MICROBIAL FUEL CELLS APPLIED TO THE METABOLICALLY - BASED DETECTION OF EXTRATERRESTRIAL LIFE

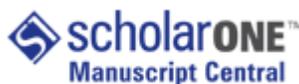

# MICROBIAL FUEL CELLS APPLIED TO THE METABOLICALLY - BASED DETECTION OF EXTRATERRESTRIAL LIFE

**Ximena C. Abrevaya[1], Pablo J. D. Mauas[1], Eduardo Cortón[2]**

[1]Instituto de Astronomía y Física del Espacio (IAFE), UBA-CONICET. Buenos Aires, Argentina.

[2]Grupo de Biosensores y Bioanálisis, Departamento de Química Biológica, Facultad de Ciencias Exactas y Naturales, UBA. Buenos Aires, Argentina.

Correspondence

Ximena C. Abrevaya. Address: CC 67, Suc. 28, 1428 Buenos Aires, Argentina. Phone: (+54)-11-4788-1916 ext.: 105. Fax: (+54)-11-4786-8114. E-mail: abrevaya@iafe.uba.ar

Running title: Microbial Fuel Cells and life detection

ABSTRACT

Since the 1970's, when the Viking spacecrafts carried out experiments aimed to the detection of microbial metabolism on the surface of Mars, the search for non-specific methods to detect life *in situ* has been one of the goals of astrobiology. It is usually required that the methodology can detect life independently from its composition or form, and that the chosen biological signature points to a feature common to all living systems, as the presence of metabolism.

In this paper we evaluate the use of Microbial Fuel Cells (MFCs) for the detection of microbial life *in situ*. MFCs are electrochemical devices originally developed as power electrical sources, and can be described as fuel cells in which the anode is submerged in a medium that contains microorganisms. These microorganisms, as part of their metabolic process, oxidize organic material releasing electrons that contribute to the electric current, which is therefore proportional to metabolic and other redox processes.

We show that power and current density values measured in MFCs using microorganism cultures or soil samples in the anode are much larger than those obtained using a medium free of microorganisms or sterilized soil samples, respectively. In particular, we found that this is true for extremophiles, usually proposed to live in extraterrestrial environments. Therefore, our results show that MFCs have the potential to be used to detect microbial life *in situ*.

INTRODUCTION

The search for life in other planets has been one of the main objectives of different space missions for the last thirty years. Many astrobiological investigations aim to find methods which could be applied to detect life or evidence of life outside the Earth. The first set of biological extraterrestrial experiments was that of the Viking Mission in 1976, which searched evidence of metabolism as a signal of putative microbial life. Three life-detection tests were performed inside the Viking landers, looking for production or absorption of volatile compounds in samples of Martian soil: the Pyrolitic Release (PR) experiment, that tested the incorporation of radioactive $CO_2$ and CO into the organic fraction of a soil sample, the Labeled Release (LR) experiment, that used radiorespirometry for the detection of metabolism or growth using radioactive nutrients labeled with $^{14}C$, and the Gas Exchange (GEX) experiment, that was particularly designed to detect signals of metabolism measuring compositional changes in the atmosphere above the soil sample after the addition of a nutrient medium. Most of these experiments searched for activity driven by a process of chemical or even biological origin. In fact, the PR experiment found that a small proportion of gas was converted into organic material, which was not reproduced in the control experiment; the GEX experiment detected increases in several biogenic gases above the soil, and the LR experiment identified the presence of radioactive gas, absent in the control experiment (Klein *et al*., 1976; Klein, 1977).

However, the gas chromatography-mass spectrometry chemical analysis which was additionally performed, failed to **find** traces of organic compounds in the Martian soil samples, a fact which would be indicative of the absence of life. The overall general consensus was that positive results obtained from biological tests were due to the presence of strong oxidizing agents on the Martian surface rather than activity from biological origin. Various investigations reviewed the controversy that followed these findings (Klein, 1977; Klein, 1979; Ponnamperuma *et al*., 1977; Owen, 1979; Navarro Gonzalez *et al*., 2006; Biemann, 2007).

At present, current approaches use several new techniques and employ a wide variety of biomarkers for the detection of different biosignatures (Kounaves *et al.*, 2002; Schweitzer *et al.*, 2005; Sims *et al.*, 2005; Parnell *et al.,* 2007; Suo *et al.*, 2007; Tang, 2007).

The discovery of extremophiles and the capacity of microorganisms to adapt to different physicochemical conditions exploiting versatile metabolic pathways, focus the research in developing new methods for *in situ* search and detection of microbial activity in extreme environments on Earth or in extraterrestrial settings. These include, among others, the use of microelectrodes for measurements of pH, minerals or natural gradients of gases, which provide information about metabolic activity (Horneck, 2000).

Although there is no general consensus about the definition of life, it is usually assumed that life should have signatures which could be detected and quantified to a certain degree. Even if most attempts are directed to detect life as we know it on Earth, the methods to detect bio-signatures should be the least earth-centric possible (Conrad and Nealson, 2001). To produce reliable results, it is usually required that the search method be repeatable and sensitive enough to detect life (Conrad and Nealson, 2001; Nealson *et al.*, 2002; Schweitzer *et al.*, 2005 and Kounaves *et al.*, 2002) and can:

i) Distinguish between biotic and abiotic processes.

ii) Be applied outside the Earth.

iii) Perform a fairly rapid scan of a large spatial region.

iv) Detect life independently from its composition or form, looking for general attributes.

The biosignature to be measured should be a distinguishable and common feature of all living system (Nealson *et al*., 2002). In particular, one of the best's signatures of life is metabolic activity, which is common to all organisms. Through metabolism organisms have the ability to convert energy from the environment into biologically usable energy. This process involves chemical oxidation-reduction (redox) reactions in which an electron donor is oxidized, and the released electrons are accepted by an electron acceptor which become reduced. Also, electron flow across a cellular membrane is used to generate an electrochemical gradient that will end up in the synthesis of ATP, the main molecular unit of chemical energy storage and transfer in the cell. The biological-driven redox gradients and electron flow involved in these reactions can be detectable and measurable. Regardless of the high metabolic diversity that we can find in known terrestrial microorganisms, all of them use the same basic process, which is present since the origin of life on Earth (Guzman and Martin, 2009; Srinivasan and Morowitz, 2009).

The capability of organisms to carry out this process is the basis of Microbial Fuel Cells (MFCs), electrochemical devices originally designed for electricity production. They work in a similar way to a battery and are usually composed by a cathode and anode separated by a cation exchange membrane. An MFC could be described as a fuel cell in which the anode is submerged in a medium (the anolyte) that contains microorganisms which, as part of their metabolic process, oxidize organic material releasing electrons and protons. The electrons are captured by the anode, built of a corrosion-resistant conductor material, and travel through an external conductor, reaching the cathode and closing the circuit. In the cathode, the reduction reactions can be achieved by contact with an oxidizing agent, like oxygen or ferricyanide among others. The protons liberated in the proximity of the anode could reach the cathode through a polymer electrolyte membrane (such as Nafion®) to complete redox reactions. This electron flow, proportional to metabolic and other redox processes in the anode region, produce a current which can be easily measured if an adequate resistor is incorporated in the electric circuit. In this way MFCs couples the metabolism of a microorganism to an electrical circuit.

The first description of MFCs is due to Potter (1911), who worked with cultures of yeast and *E. coli*. Later, they were rediscovered by Bennetto (1984; see also Allen and Bennetto 1993). At present, there is much work done with MFCs, with multiple combinations of electrode material and microorganisms (see the reviews by Rabaey *et al.* 2005; Bullen *et al.* 2006; Davis and Higson, 2007). In particular, Miller and Oremland (2008) used extremophile bacteria as biocatalysts, and suggested the use of MFCs as life detectors.

Since extremophiles live in habitats with extreme physicochemical conditions, which are considered as analogs of extraterrestrial environments, they are usually proposed as possible inhabitants of other planets (Cavicchioli, 2002). In particular, the haloarchaea are extremophile microorganisms that inhabit in **hypersaline** environments (3–5 M NaCl) such as salt lakes, marine salterns, or salt ponds. They have a special importance in astrobiology because they were also found entrapped in salt deposits in ancient evaporitic rocks as halites (Grant *et al*., 1998; Grant, 2004; Fendrihan *et al*., 2006). Since the same structures were indentified in SNC martian meteorites (Gooding, 1992; Zolensky *et al*., 1999; Whitby *et al*., 2000; Stan Lotter *et al*., 2004), it is thought that they could be present on Mars or on other planets that harbor saline environments (Sims *et al*., 1997; Grant, 2004; Fendrihan *et al*., 2006).

In the present paper we investigate the capacity of MFCs to detect the presence of metabolism, and its use to search *in situ* for extraterrestrial life, and we show that this method satisfies all the requirements previously mentioned. In particular, we report new experimental results using a culture of **an archaeon,** which was not previously employed as biocatalyst in MFCs. We also used a mixed community of microorganisms from humus soil. Therefore, our conclusions were obtained based on representatives of the three domains of life, Archaea, Bacteria and Eukarya, which are taxonomically and metabolically diverse.

MATERIALS AND METHODS:

**<u>MFC design and operation</u>:**

MFCs used in the present work (Fig. 1) were made of polystyrene cylinders containing at the bottom a Nafion® membrane (DuPont, Wilmington, DE) which allows, at the same time, separation from the outside and proton exchange between the inside and the outside media, the one to be analyzed for redox activity. Inside this tube is the cathodic compartment, the catholyte, that contains a solution made of ferricyanide (8.4 g $L^{-1}$) dissolved in an appropriate buffer solution, according to the sample to be measured. Outside the tube, the anodic compartment behaves as an anolyte, containing a sample of the microorganisms to be studied (culture or soil), which in the control experiments were replaced by sterilized culture or soil.

The cathode and anode were made of Toray® carbon paper (TGP-H-120, density, 0.45 g $cm^{-3}$, porosity 78%, Toray Industries) for microbial culture experiments and graphite reinforced with carbon (0.7 HB pencil lead, Faber Castell, Brasil) for soil experiments. Both electrodes were connected to insulated copper wire with silver epoxy, and covered thoroughly with insulating epoxy, to avoid spurious currents that might arise from copper oxidation. The total area of the electrodes was 5.5 $cm^2$ for carbon paper and 3.76 $cm^2$ for graphite reinforced with carbon (anode and cathode areas are identical), and the area of the Nafion® membrane was 0.38 $mm^2$.

To study the MFC behavior the potential (E, Volts) was measured at open circuit using a Fluke 289 digital tester (Fluke Inc., Everett, WA), and using different resistances ranging from 98$\Omega$ to 100 k$\Omega$ as an external load. The electrical current (I, Ampere) was calculated according to Ohm's Law

$$E = I.R$$

and power (P, Watt) was calculated according to

$$P = I.E$$

where R is the resistance ($\Omega$, Ohm). Power and electrical current were normalized using the electrode area and were expressed as power and current density, (p) in Watt cm$^{-2}$ and (j) in Ampere cm$^{-2}$ .

**Microbial culture experiments**

We considered two experimental groups: sterilized microorganism cultures (autoclaved at 121ºC, 45 min) and non-sterilized microorganism cultures. Both were placed in the anolyte compartment. To test the capacity of MFCs to detect life, we used two different microorganisms:

*Saccharomyces cerevisiae* (Baker's yeast) is a facultative anaerobic unicellular fungus which belongs to the Eukarya domain.  Industrial *S. cerevisiae* (dry active yeast, from Calsa S.A., Argentina) was prepared adding 3 g of the dry powder to 60 mL of phosphate buffer 100 mM, adjusting pH to 7. Glucose 5 g $L^{-1}$ was added as a carbon source, and aliquots of 60 mL for sterilized and non-sterilized groups were placed as anolytes and incubated in a thermostatic bath at $37 \pm 1°C$ for the measurements.

*Natrialba magadii* is a prokaryotic microorganism which belongs to the Archaea domain and is a heterotrophic aerobic member of the family *halobacteriaceae*. *N. magadii* was originally isolated from Magadii Lake, in Kenya, Africa (Tindall *et al*., 1984). It is an extremophile haloalkalophile microorganism that lives in 3.5-4.0 M NaCl and pH values between 9 and 11 (optimum range). *N. magadii* (ATCC 43099) was grown aerobically at 37°C. Growth medium contains (g $L^{-1}$): yeast extract (10); NaCl (200); $Na_2CO_3$ (18.5); Sodium citrate (3); KCl (2); $MgSO_4.4H_2O$ (1); $MnCl_2.4H_2O$ ($3.6x10^{-4}$); $FeSO_4.7H_2O$ ($5x10^{-3}$); pH was adjusted to 10.

*N. magadii* liquid culture (at optical density around 0.5, exponential phase) was divided in aliquots of 36 ml. The aliquots for sterilized and non sterilized groups were centrifuged 5 min at 45 rpm. Supernatant was discarded and pellet was resuspended in 36ml of fresh medium, then collected in a flask and diluted to reach 60 ml and used as anolytes.

A resistance of 4600 $\Omega$ was coupled to the circuit. After 12 hours, this resistance was disconnected and measurements were taken two hours later. The experiments were carried out at $37 \pm 1°C$ in a thermostatic bath.

The catholyte was composed of ferricyanide (8.4 g $L^{-1}$) dissolved in a phosphate buffer solution (100 mM, pH=7) for *S. cerevisiae* experiments and of saline solution (NaCl (200); $Na_2CO_3$ (18.5); Sodium citrate (3); KCl (2); $MgSO_4.4H_2O$ (1); $MnCl_2.4H_2O$ ($3.6x10^{-4}$); $FeSO_4.7H_2O$ ($5x10^{-3}$), pH=10) for *N. magadii*.

## Soil experiments

To test the MFC capacity to detect life in soil samples we used humus-rich (about 5 %) topsoil, commercially available. The sample was divided in two sub-samples of 70g, sterile (control) and non-sterile, both in duplicate. For control, sterilized soil was obtained by autoclaving at 121ºC in two cycles of 1h 45 min and 45 min, separated by 24 hours. Then, 40 ml of distilled water were added to each sample and voltage was measured at the beginning of the experiment and every 24 hours during 144 hours. Each day an external resistance of 4600 $\Omega$ was coupled to the circuit. The resistance was disconnected two hours before each measurement was done. Catholyte was phosphate buffer (100 mM, pH=7) containing ferricyanide. The experiments were carried out at room temperature (27ºC $\pm$ 1ºC).

## RESULTS

Culture experiments

In Fig. 2 and 3, we show the results obtained for the *S. cerevisiae* and *N. magadii* cultures, respectively, which are also summarized in Table 1. In the figures we show potential and power density as a function of current density. It can be seen that in both experiments power densities were larger when microorganisms were present in the anode than in the corresponding sterilized samples. On the other hand, in the experiments with microorganisms the maximum power densities occur at higher current densities than in the control experiments (Table 1).

Soil experiments

In Fig. 4, we plot the evolution of potential with time, with the system loaded with a fixed resistance (4600Ω). It can be seen that after around 48 hours the potential in the non-sterile samples increased, as a signal of the presence of microorganisms. On the other hand, the sterile samples showed no increment of the potential. In Fig. 5, we show the potentials, power and current densities obtained for the soil experiments, which are also summarized in Table 1. Here, we show potential and power density as a function of current density, measured 114 hours after the beginning of the experiments. Also in this case, in the experiments with microorganisms the power densities are much larger, and the maximum occurs at higher current density than in the sterile soil.

DISCUSSION

Our experiments show that power and current densities are significantly larger when microorganisms are present in the samples. This is due to the coupling of the microorganism's metabolism to the electrical circuit. Therefore, our MFCs results demonstrate that this method can detect the presence of microbial life, satisfying the requisites mentioned in the Introduction, and could be used for the *in situ* search of extraterrestrial life.

In principle, there are no impediments to apply this method outside the Earth, and it can be applied to liquid samples, or after a short period of time in soil experiments**.** In this case, a rapid scan of a large spatial region could be performed taking different soil samples and performing several experiments at the same time. Several **studies** show that MFCs can be employed in different experimental conditions, such as pH values ranging from acid to alkaline (Liu *et al.*, 2010; Raghavulu *et al.*, 2008). Temperature working values were found in a range from 5 to 60ºC (Mathis *et al.*, 2007; Scott *et al.*, 2008). Up to now, salt concentrations used in MFCs were in the range 0.1 to 0.4 M NaCl (Huang *et al.*, 2010; Liu *et al.*, 2005). In the present work we tested a concentration 10 times larger than previously **documented. Therefore,** we demonstrate that the range of ionic strength in which MFCs can operate could be expanded to work in hypersaline environments and for the detection of extreme halophiles.

Aditionally it is important to note that, for the first time, we used MFCs with archaea, which constitute the vast majority of the extremophiles and we found that the values obtained in this case were larger than the other microorganisms. In particular, *N. magadii* is a halophile archaeon, and halophiles are usually proposed as "exophiles", due to their capacity to survive to extreme physicochemical conditions: not only high salt concentrations, but also high UV or ionizing radiation doses, low oxygen levels, extreme temperatures and pH values (DasSarma, 2006).

Our approach in this work is similar to the one used to test on the ground the biological experiments before the launch of the Viking mission. In particular, the LR experiments were performed on pure cultures of microorganisms and soil samples (Levin and Straat, 1976), as we did to explore the ability of MFCs to detect *in situ* microbiological life. Furthermore, our experiments involved different kinds of microorganisms, representatives of the three domains of life, Archaea, Bacteria and Eukarya, which are taxonomically and metabolically diverse, and in all cases the results were positive. Therefore, the method is able to detect different kind of metabolisms, and allows the detection of life independently of its form, which is part of the main requirements for this kind of tests.

However, in contrast to the Viking experiments, our methodology does not require the existence of a carbon based life, and it is, therefore, a more general approach to detect life independently from its composition.

ACKNOWLEDGEMENTS

The authors would like to thank Diego De Gennaro for the MFC scheme, and the two anonymous reviewers, who helped improve the paper.

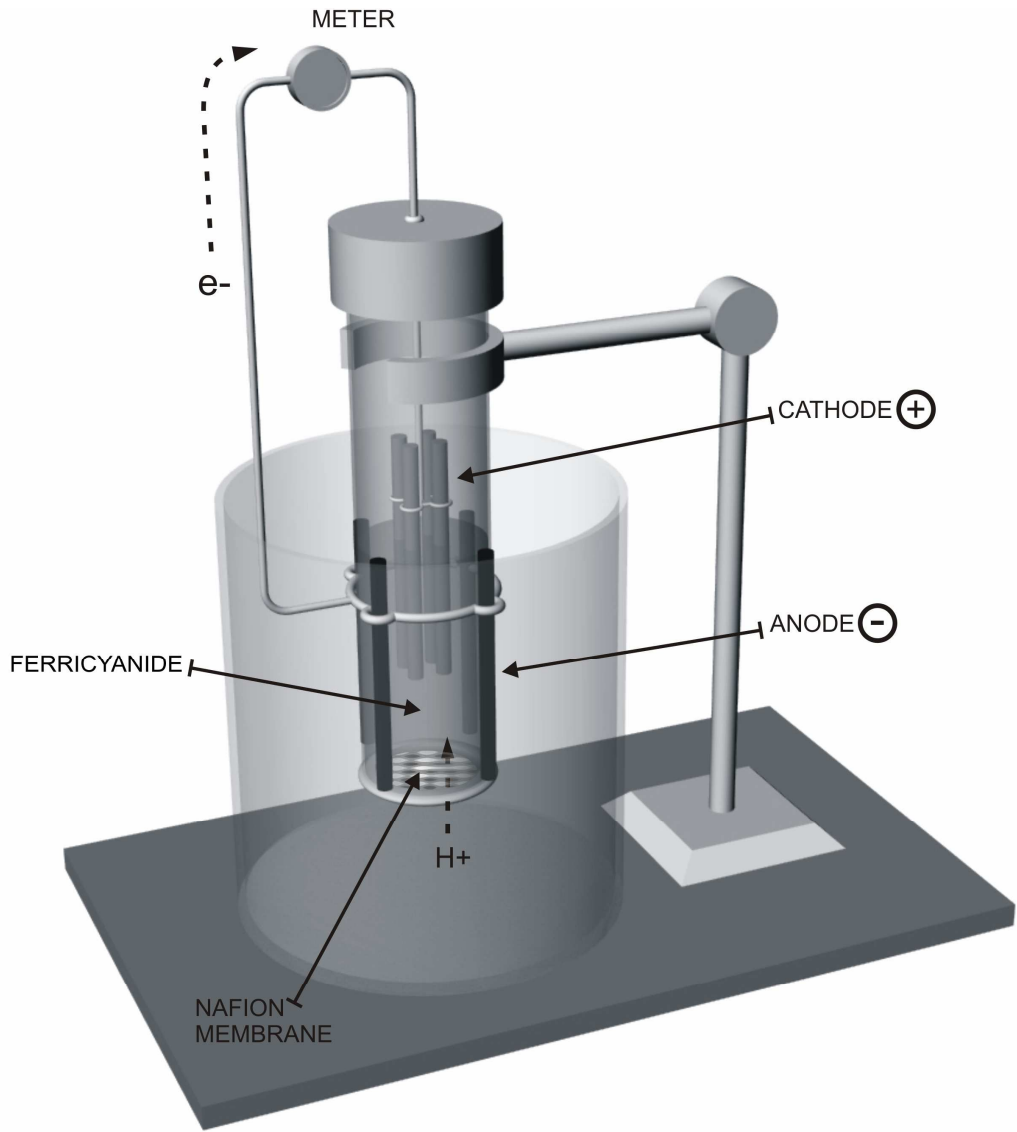

Figure 1. Schematic representation of submersible MFC used to perform the experiments.

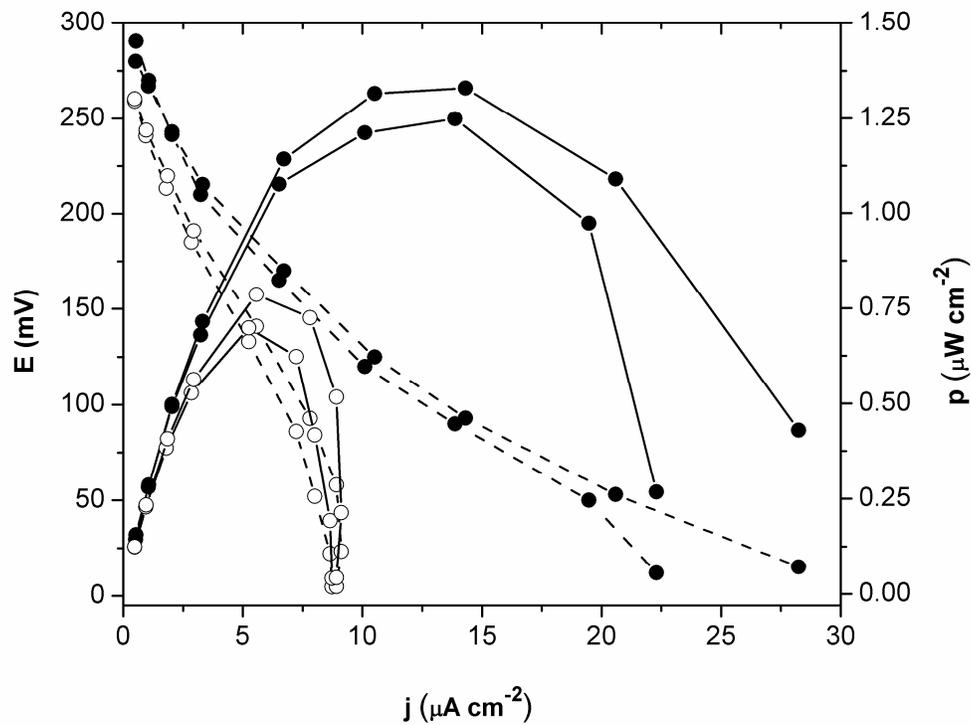

FIG. 2. MFC power density (solid lines, right scale) and potential (dashed lines, left scale) as a function of current density for *S. cerevisiae*. **Experiments were performed in duplicate.** Open circles: sterilized culture. Solid circles: non-sterilized culture.

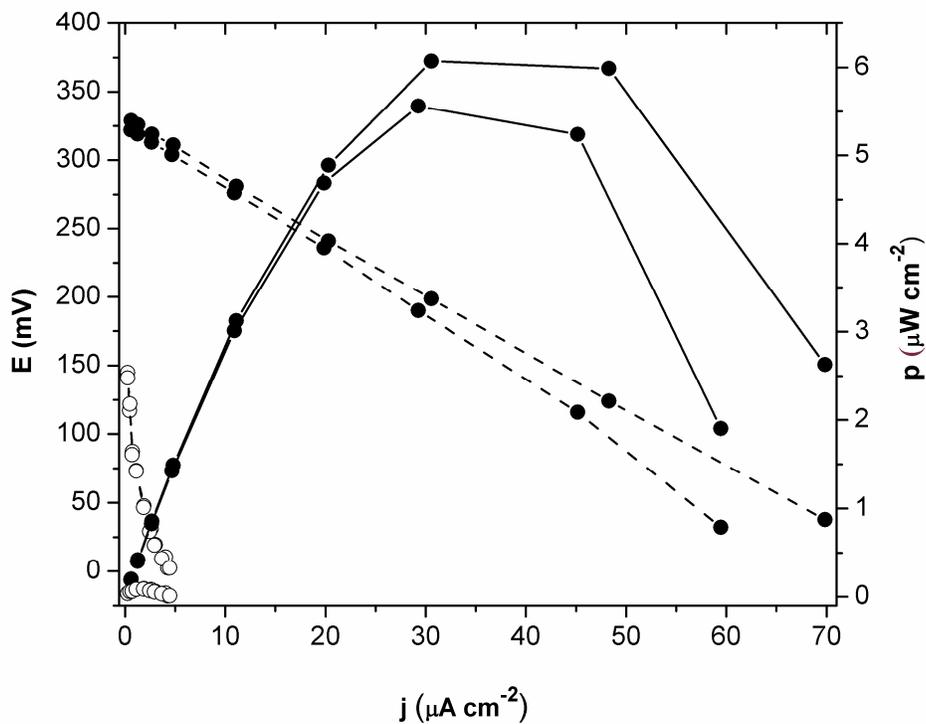

FIG. 3. MFC power density (solid lines, right scale) and potential (dashed lines, left scale) as a function of current density for *N. magadii*. **Experiments performed in duplicate are shown.** Open circles: sterilized culture. Solid circles: non-sterilized culture.

1
2
3
4
5
6
7
8
9
10
11
12
13
14
15
16
17
18
19
20
21
22
23
24
25
26
27
28
29
30
31
32
33
34
35
36
37
38
39
40
41
42
43
44
45
46
47
48
49
50
51
52
53
54
55
56
57
58
59
60

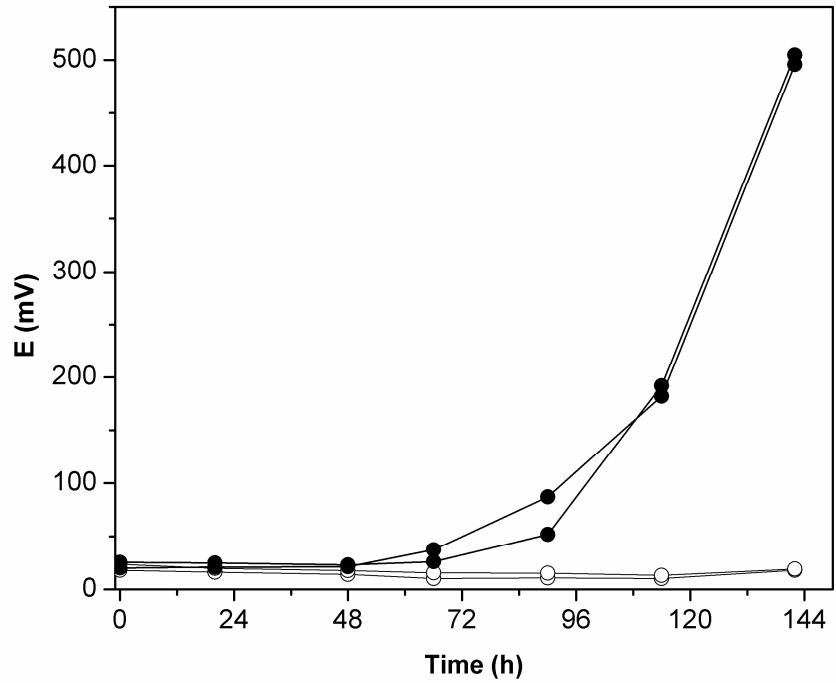

FIG. 4. MFC potential as a function of time, R = 4600Ω for soil experiments.

**Experiments in duplicate for sterile (open circles) and non-sterile samples**

**(solid circles) are plotted.**

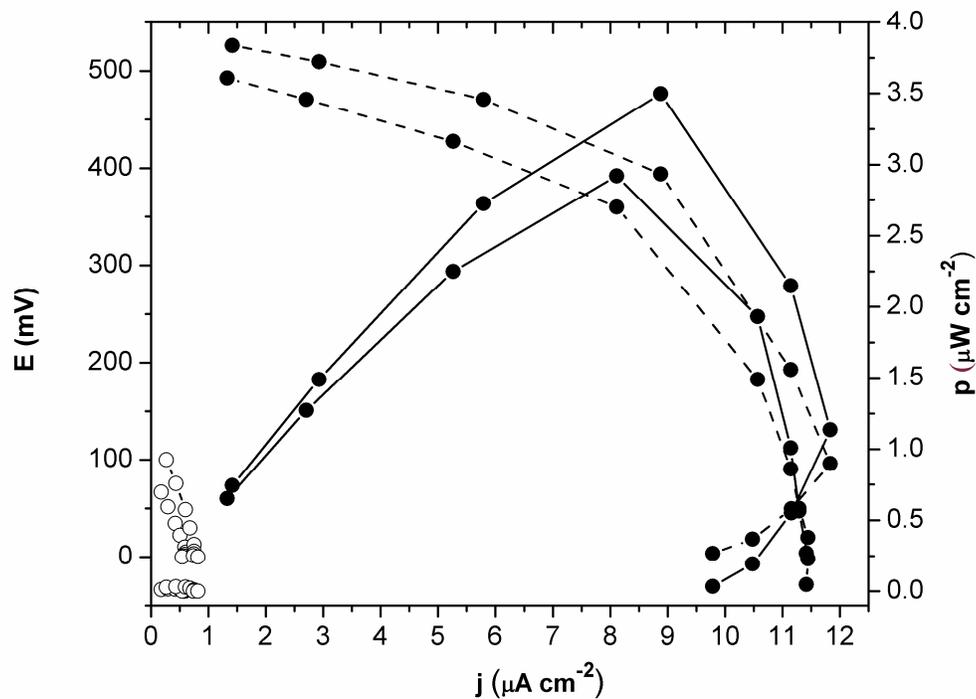

FIG. 5. MFC power density (solid lines, right scale) and potential (dashed lines, left scale) as a function of current density in soil experiments **performed in duplicate.** Open circles: sterilized soil. Solid circles: non-sterilized soil.

Table 1. Summary of different values obtained from culture and soil experiments. In all cases we list the mean value of the duplicate experiments ± half the difference between them.

| Sample/Experiment | Max. Power density ($\mu$W cm$^{-2}$) | Current Density at max. power density ($\mu$A cm$^{-2}$) | Potential at open circuit (mV) |
|---|---|---|---|
| **Pure culture experiments** | | | |
| *S. cerevisiae* (sterile) | 0.74 ± 0.04 | 5.41 ± 0.16 | 277.0 ± 4.8 |
| *S. cerevisiae* (non-sterile) | 1.28 ± 0.03 | 14.00 ± 0.22 | 305.5 ± 0.5 |
| *N. magadii* (sterile) | 0.090 ± 0.002 | 1.860 ± 0.021 | 256.5 ± 2.5 |
| *N. magadii* (non-sterile) | 5.82 ± 0.25 | 30.00 ± 0.66 | 333.5 ± 8.5 |
| **Soil experiment** | | | |
| Sterile soil (t = 114h) | 0.024 ± 0.085 | 0.36 ± 0.07 | 11.60 ± 1.40 |
| Non-sterile soil (t = 114h) | 3.21 ± 0.29 | 8.50 ± 0.38 | 187.0 ± 5.01 |

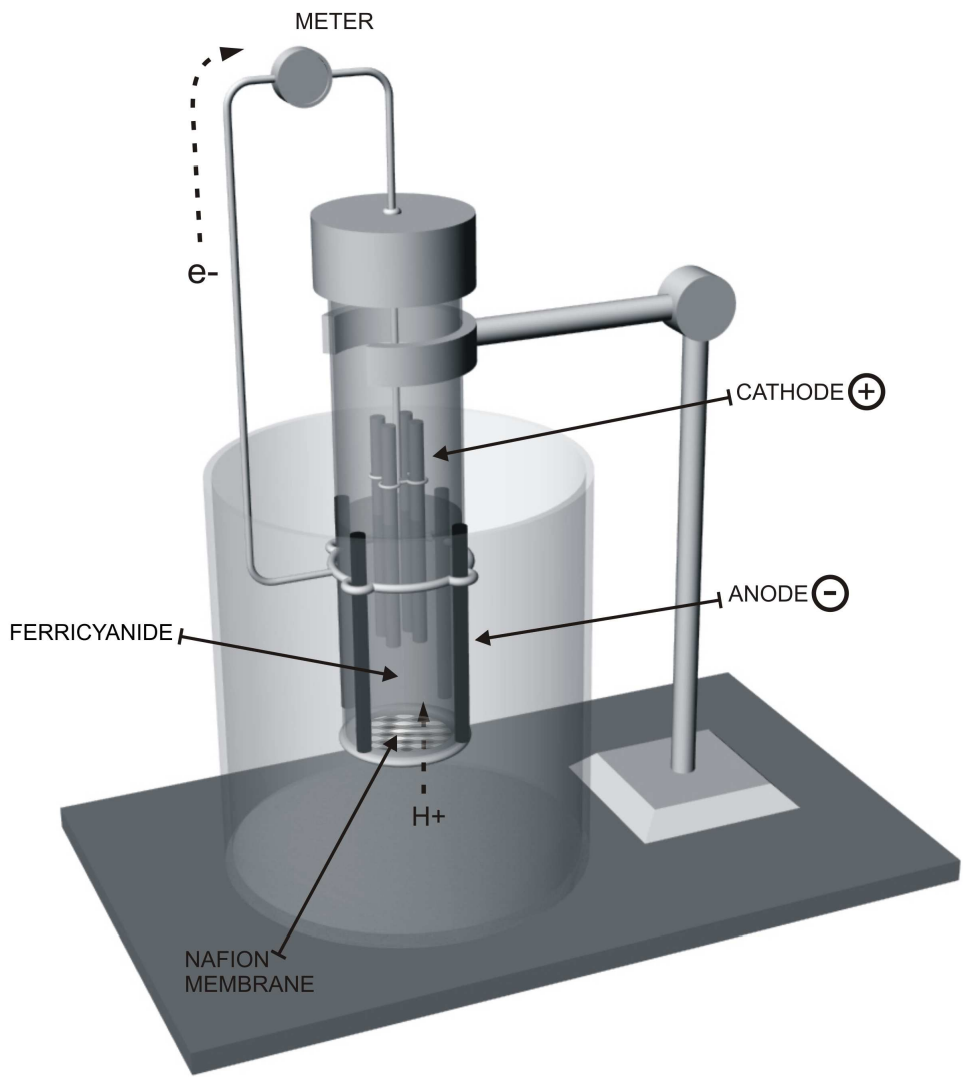

Figure 1. Schematic representation of submersible MFC used to perform the experiments.

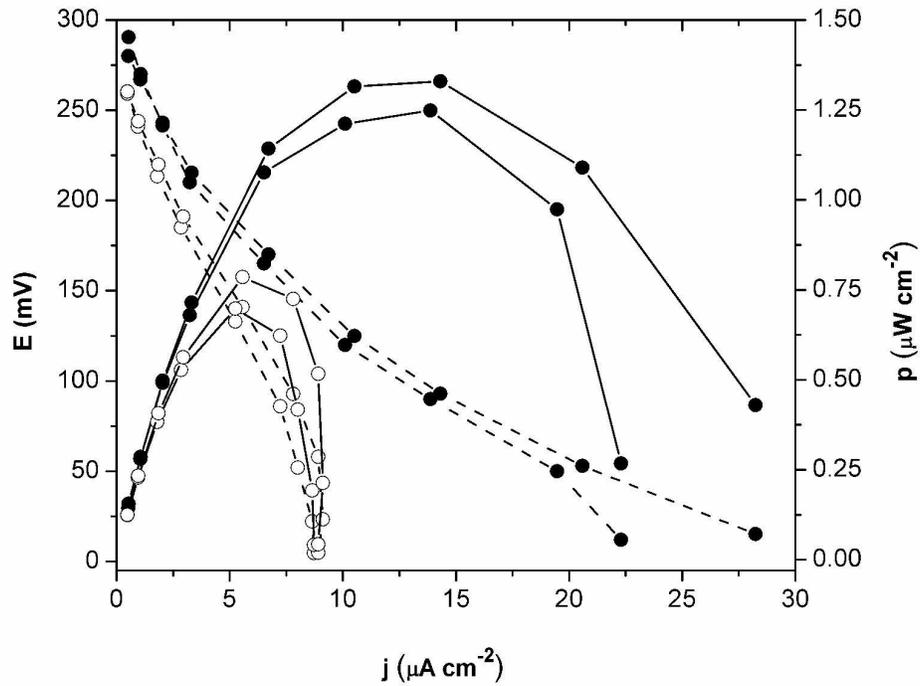

FIG. 2. MFC power density (solid lines, right scale) and potential (dashed lines, left scale) as a function of current density for *S. cerevisiae*. Experiments were performed in duplicate. Open circles: sterilized culture. Solid circles: non-sterilized culture.
279x215mm (300 x 300 DPI)

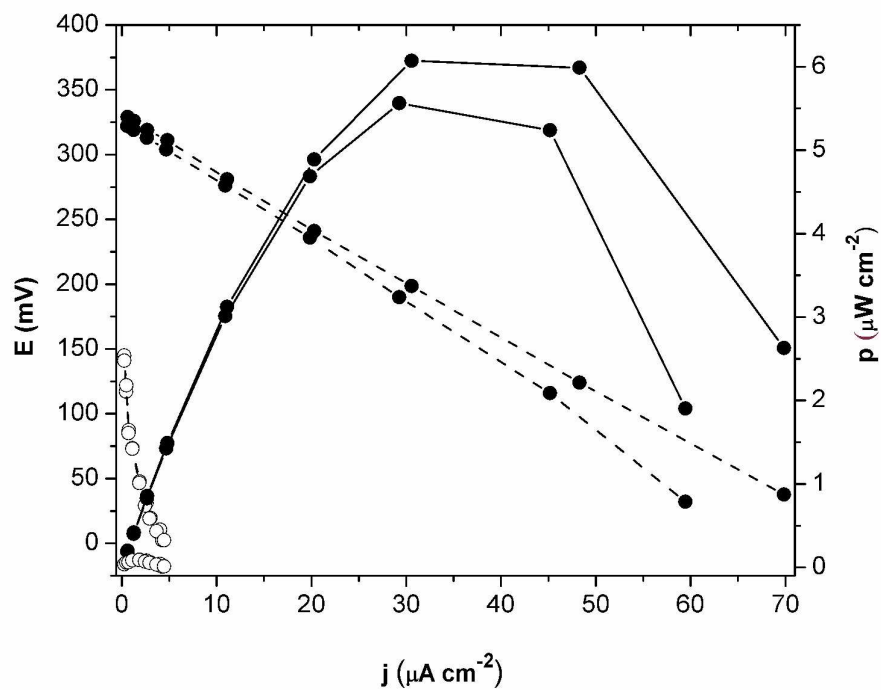

FIG. 3. MFC power density (solid lines, right scale) and potential (dashed lines, left scale) as a function of current density for *N. magadii*. Experiments performed in duplicate are shown. Open circles: sterilized culture. Solid circles: non-sterilized culture.

279x215mm (300 x 300 DPI)

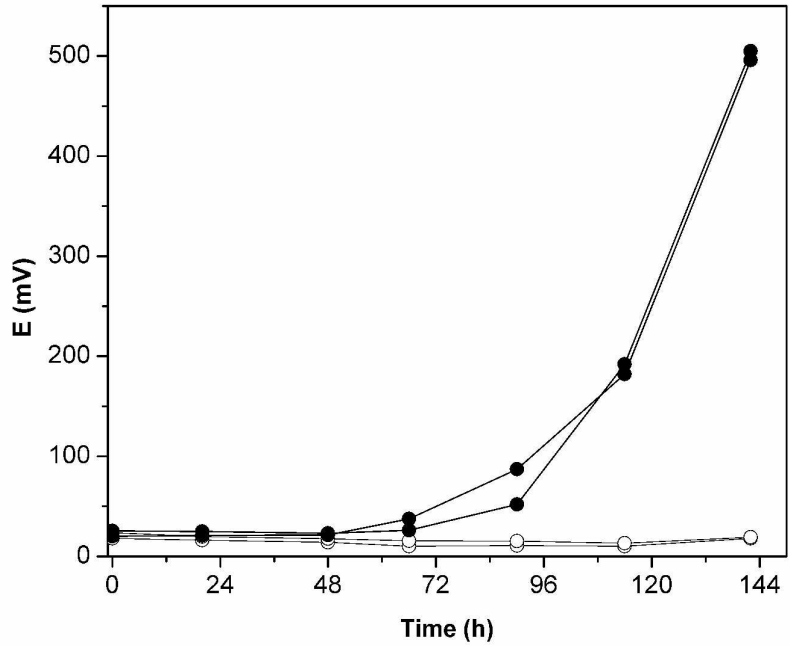

FIG. 4. MFC potential as a function of time, R=4600Ω for soil experiments. Experiments in duplicate for sterile (open circles) and non-sterile samples (solid circles) are plotted.
279x215mm (300 x 300 DPI)

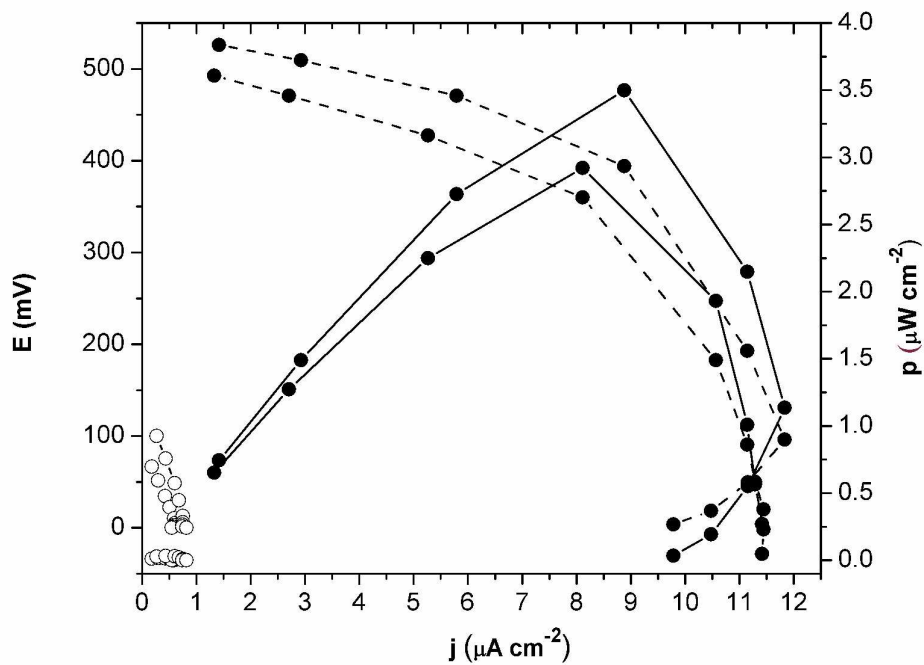

FIG. 5. MFC power density (solid lines, right scale) and potential (dashed lines, left scale) as a function of current density in soil experiments performed in duplicate. Open circles: sterilized soil. Solid circles: non-sterilized soil.
279x215mm (300 x 300 DPI)